\documentclass[aps,prd,reprint,groupedaddress,noeprint,nolongbibliography]{revtex4-2}

\usepackage{amssymb}
\usepackage{graphicx}
\usepackage{color}

\newcommand{\mnras}{Mon. Not. R. Astron. Soc.}

\newcommand{\be}{\begin{equation}}
\newcommand{\ee}{\end{equation}}
\newcommand{\ba}{\begin{eqnarray}}
\newcommand{\ea}{\end{eqnarray}}
\newcommand{\lp}{\left(}
\newcommand{\rp}{\right)}
\newcommand{\lb}{\left[}
\newcommand{\rb}{\right]}

\newcommand{\nus}{f_\mathrm{spin}}
\newcommand{\Msun}{M_{\rm Sun}}
\newcommand{\nugw}{f}
\newcommand{\mchirp}{{\cal M}}
\newcommand{\Omegaorb}{\Omega_{\rm orb}}
\newcommand{\rorb}{a}
\newcommand{\Eorb}{E_{\rm orb}}
\newcommand{\torb}{t_{\rm orb}}
\newcommand{\tinspiral}{t_D}
\newcommand{\Delphi}{\Delta\Phi}
\newcommand{\omegamode}{\omega_\alpha}
\newcommand{\omegamodend}{\hat{\omega}_\alpha}
\newcommand{\omegazero}{\omega_1}
\newcommand{\Qmode}{Q_\alpha}
\newcommand{\ximode}{\xi_\alpha}
\newcommand{\Apg}{A_0}
\newcommand{\Npg}{N_0}
\newcommand{\npg}{n_0}
\newcommand{\nusat}{f_0}
\newcommand{\nuref}{f_{\rm ref}}

\begin{document}

\title{New dynamical tide constraints from current and future \\ gravitational wave detections of inspiralling neutron stars}

\author{Wynn C. G. Ho}
\email[]{wynnho@slac.stanford.edu}
\affiliation{Department of Physics and Astronomy, Haverford College, 370 Lancaster Avenue, Haverford, PA, 19041, USA}
\author{Nils Andersson}
\affiliation{Mathematical Sciences and STAG Research Centre, University of Southampton, Southampton SO17 1BJ, UK}

\date{\today}

\begin{abstract}
Previous theoretical works using the pre-merger orbital evolution of
coalescing neutron stars to constrain properties of dense nuclear
matter assume a gravitational wave phase uncertainty of a few radians,
or about a half cycle.
However, recent studies of the signal from GW170817 and
next generation detector sensitivities indicate actual phase uncertainties
at least twenty times better.
Using these refined estimates, we show that future observations
of nearby sources like GW170817 may be
able to reveal neutron star properties beyond just radius and
tidal deformability,
such as the matter composition and/or presence of a superfluid
inside neutron stars, via tidal excitation of g-mode oscillations.
Data from GW170817 already limits the amount of orbital energy that
is transferred to the neutron star to $<2\times 10^{47}\mbox{ erg}$
and the g-mode tidal coupling to $\Qmode<10^{-3}$ at 50~Hz
($5\times 10^{48}\mbox{ erg}$ and $4\times 10^{-3}$ at 200~Hz), and future
observations and detectors will greatly improve upon these constraints.
In addition, analysis using general parameterization models that have
been applied to the so-called p-g instability show that the instability
already appears to be restricted to regimes where the mechanism is
likely to be inconsequential;
in particular, we show that the number of unstable modes is $\ll 100$ at
$\lesssim 100\mbox{ Hz}$,
and next generation detectors will essentially rule out this mechanism
(assuming that the instability remains undetected).
Finally, we illustrate that measurements of tidal excitation of
r-mode oscillations in nearby rapidly rotating neutron stars are within reach
of current detectors and note that even non-detections will
limit the inferred inspiralling neutron star spin rate to $< 20\mbox{ Hz}$,
which will be useful when determining other parameters
such as neutron star mass and tidal deformability.
\end{abstract}

\maketitle

\section{Introduction\label{sec:intro}}

Measurements of gravitational waves (GWs) from coalescing
neutron star (NS) systems provide invaluable insights into the
dense matter that comprises the interior of these stars.
As is now well-known, the detection of GWs from the inspiral and merger
of GW170817 enabled new constraints on the nuclear equation-of-state
through determination of the radius and tidal deformability of each
NS \cite{abbottetal17,abbottetal18}.
With advancements in GW detector sensitivities, it may be possible
to obtain measurements from future discoveries that provide constraints
which go beyond just bulk NS properties.
One example, which is the subject of the current work, is dynamical
tidal excitation of NS oscillation modes.
Such a process can occur as the orbital frequency increases during
the binary inspiral and comes into resonance with the natural oscillation
frequencies of the NS.
As a result, energy can be transferred
between the orbit and the stellar oscillation, which causes the
inspiral to occur more rapidly (or more slowly) and creates a
phase shift in the GW signal.
By measuring the phase shift and
GW frequency at which the phase shift occurs, it may then be possible to
infer physical properties of the specific oscillation
mode, for example, the particle fractions in the NS or the superfluid
state of the star's core.

In this work, we will only be concerned with mode, orbital, and GW frequencies
below a few hundred Hz, i.e., well before NS merger,
where differences in current GW waveform models are much smaller than the
estimated data uncertainties \cite{read23}.
As such, the strongest coupling between the gravitational tidal
potential and an oscillation mode (characterized by $\alpha=nlm$,
where $n$ is the number of radial nodes of the mode displacement eigenfunction
$\ximode$ and $l$ and $m$ are indices of the spherical harmonics $Y_{lm}$)
will be through g-modes (see Section~\ref{sec:discuss} for r-modes).
These g-modes are fluid oscillations whose restoring force is
buoyancy caused by, for example, changes in the proton to neutron
or muon to electron fraction as a function of density, with the
latter being important when neutrons are superfluid
\cite{kantorgusakov14,anderssonho18}.
Furthermore, by considering only the leading order tidal quadrupole
and primarily non-rotating stars, one only needs to consider the
quadrupole modes with $l=2$.
The strength of the tide-mode coupling is determined by the (dimensionless)
overlap integral
$\Qmode=(1/MR^l)\int d^3x\,\rho\ximode^\ast\cdot\nabla(r^lY_{lm})$,
where $M$ and $R$ are stellar mass and radius, respectively,
and $\rho$ is mass density \cite{pressteukolsky77}.
Early works found the largest $\Qmode$ ($\sim10^{-3}$) for low $n$ g-modes
\cite{lai94,kokkotasschafer95} and that these values of $\Qmode$
produce phase shifts $\Delphi\sim10^{-3}\mbox{ rad}$
\cite{lai94,reiseneggergoldreich94},
which were thought to be too small to be detectable given estimates of
GW detector uncertainties at that time (see below).

A related phenomena which we also examine here is the p-g instability,
where instead of tidal resonances with specific oscillation modes
during the inspiral, instabilities between coupled p and g-modes
supposedly become excited by the tidal potential at GW frequencies
$f$ of a few tens of Hz and the orbit
loses energy continuously to these modes throughout the remainder
of the inspiral \cite{weinbergetal13,weinberg16,essicketal16,abbottetal19}.
Because the p-g instability is not a resonant process and the
energy loss occurs over a larger frequency range, a change in
phase can gradually build up to large values during the NS inspiral, and
hence it was suggested that the instability could strongly alter the GW signal.
It was originally estimated that an effective p-g instability amplitude
$\Apg\gtrsim10^{-8}$ would produce phase shifts $\Delphi\gtrsim1\mbox{ rad}$
\cite{essicketal16}, although subsequent work in \cite{abbottetal19}
found this amplitude
to be underestimated by a factor of $\sim(4-\npg$) (cf. Section~\ref{sec:pg}),
where $\npg$ describes the frequency dependence of the p-g instability
and is of order unity.

\begin{figure}
\includegraphics[width=0.45\textwidth]{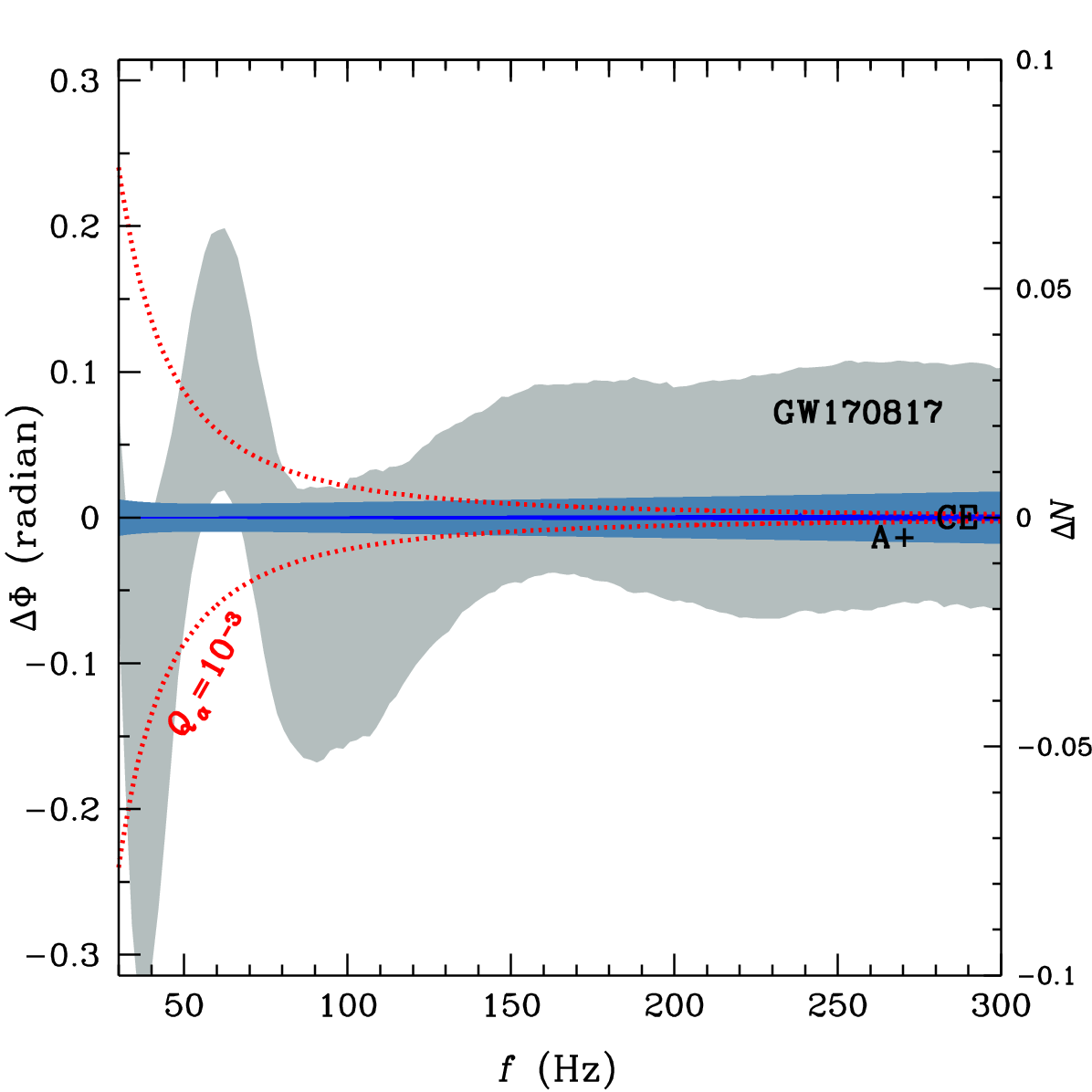}
\caption{
Phase uncertainty as a function of GW frequency $\nugw$.
Shaded regions indicate $\pm1\sigma$ deviation of $\Delphi$
from a GW170817 waveform (\cite{edelmanetal21}; see text)
and from a model waveform weighted by the sensitivity of A+ and
Cosmic Explorer (CE) \cite{read23} at a distance of 40~Mpc.
Dotted lines indicate $\pm\Delphi$ assuming (dimensionless)
overlap integral $\Qmode=10^{-3}$
and NS radius $R=12\mbox{ km}$ [see eq.~(\ref{eq:qmode})].
\label{fig:dphi}}
\end{figure}

To be able to measure tidal excitation of g-modes or the p-g instability,
the phase uncertainty of detected GWs must be smaller than the phase shift
$\Delphi$ (or change in number of orbital cycles $\Delta N=\Delphi/2\pi$)
predicted by the theory of these processes.
In the works cited above, as well as others,
the phase uncertainty was assumed to be either
$\Delphi\approx 1-3\mbox{ rad}$, based on an estimate of a detection with
signal-to-noise ratio of 10 and an approximate single detector sensitivity
\cite{cutlerflanagan94,balachandranflanagan07}, or
$\Delphi=\pi$ (or $\Delta N=0.5$), based on simple matched-filter arguments.
However, recent works quantified the level of uncertainty of measured
or measurable GWs.
Specifically in \cite{edelmanetal21}, analysis of the signal from GW170817 using
the GW waveform model \texttt{IMRPhenomPv2\_NRTidal} \cite{dietrichetal19},
which includes effects of static tidal deformabilities, find
$\Delphi\sim\pm0.1\mbox{ rad}$ (or $\Delta N\sim\pm0.03$)
at GW frequencies $\nugw<300\mbox{ Hz}$,
inclusive of calibration uncertainties.
This uncertainty is shown in Figure~\ref{fig:dphi};
note that while $\Delphi$ deviate from zero at $1\sigma$ for some frequencies,
they are consistent with zero at $2\sigma$,
as shown in Figure~12 of \cite{edelmanetal21}.
More recently, \cite{read23} compares a number of GW waveform models
and shows the uncertainty due to waveform differences, and hence
the likely best possible uncertainty at the present time, is
$\sim\pm0.02\mbox{ rad}$ for A+ and $\pm10^{-3}\mbox{ rad}$
for Cosmic Explorer (CE);
we do not consider here the Einstein Telescope (ET),
but it is expected to have similar phase uncertainties to CE.
Since these phase uncertainties are determined from comparisons
between measured/expected GW data and waveform models, they can serve as
upper limits on any effects that the waveform models do not take into
account but could be present in actual GW data, such as the influence
of dynamical tides
(see, e.g., \cite{cutlerflanagan94,maetal21,guptaetal23}).

Given the magnitudes of the phase uncertainty of current detectors and
A+ and CE, it is clear that the previously adopted level of phase shift
necessary for dynamical tide effects (see, e.g., \cite{anderssonho18})
is too conservative by at least a factor of 20.
Thus future detectors are more likely to detect these effects, which
motivates reconsideration and points to work that is needed to
maximize the science that can be extracted from future observations.
With this aim in mind, we revisit our previous analysis \cite{anderssonho18}
and re-evaluate the detectability of g-modes (in Section~\ref{sec:gmode})
and the p-g instability (in Section~\ref{sec:pg}) in inspiralling NSs
(see Section~\ref{sec:discuss} for comments on r-modes).

\vspace{-1em}
\section{Orbital energy transfer and g-mode resonance\label{sec:gmode}}
\vspace{-1em}

We do not repeat here derivations of the relevant equations,
since they are contained in many previous works, and
simply state the main relations, following closely the Newtonian orbit
calculations of \cite{lai94} (see also \cite{anderssonho18}).
An estimate of the shift in orbital phase $\Delphi$
due to energy transfer $\Delta E$ during orbital decay is
\ba
\frac{\Delphi}{2\pi} &\approx& -\frac{\tinspiral}{\torb}\frac{\Delta E}{|\Eorb|}
 \nonumber\\
&=& -\frac{5c^5}{128\pi}\frac{1}{(GM_1\Omegaorb)^{5/3}}\frac{(1+q)^{1/3}}{q}
 \frac{\Delta E}{|\Eorb|} \nonumber\\
&=& -430\,M_{1.4}^{-5/3}\!\lp\frac{1+q}{2q^3}\rp^{1/3}
 \!\!\lp\frac{\nugw}{100\mbox{ Hz}}\rp^{-5/3}\!\!\!\frac{\Delta E}{|\Eorb|},
 \label{eq:denergy}
\ea
where the orbital energy is
\ba
\Eorb &=& -\frac{GM_1M_2}{2a} \nonumber\\
&=& -1.7\times10^{52}\mbox{ erg }M_{1.4}^{5/3}q\lp\frac{2}{1+q}\rp^{1/3}
 \!\!\lp\frac{\nugw}{100\mbox{ Hz}}\rp^{2/3}
\ea
for binary masses $M_1$ and $M_2$, mass ratio $q=M_2/M_1$,
orbital separation $a$, and orbital frequency $\Omegaorb$ ($=\pi\nugw$).
We also note that the chirp mass
$\mchirp=(M_1M_2)^{3/5}/(M_1+M_2)^{1/5}=M_1[q^3/(1+q)]^{1/5}$.
The two timescales are the orbital period $\torb=2\pi/\Omegaorb$
and orbital decay timescale
\ba
\tinspiral &\equiv& \frac{\rorb}{|\dot{\rorb}|}
 = \frac{5c^5}{64G^3}\frac{\rorb^4}{M_1M_2(M_1+M_2)} \nonumber\\
&=& 8.6\mbox{ s }M_{1.4}^{-5/3}\lp\frac{1+q}{2q^3}\rp^{1/3}
 \lp\frac{f}{100\mbox{ Hz}}\rp^{-8/3}.
\ea
For simplicity, we assume $M_{1.4}=M_1/1.4\,\Msun=1$ and $q=1$
throughout the present work.

\begin{figure}
\includegraphics[width=0.45\textwidth]{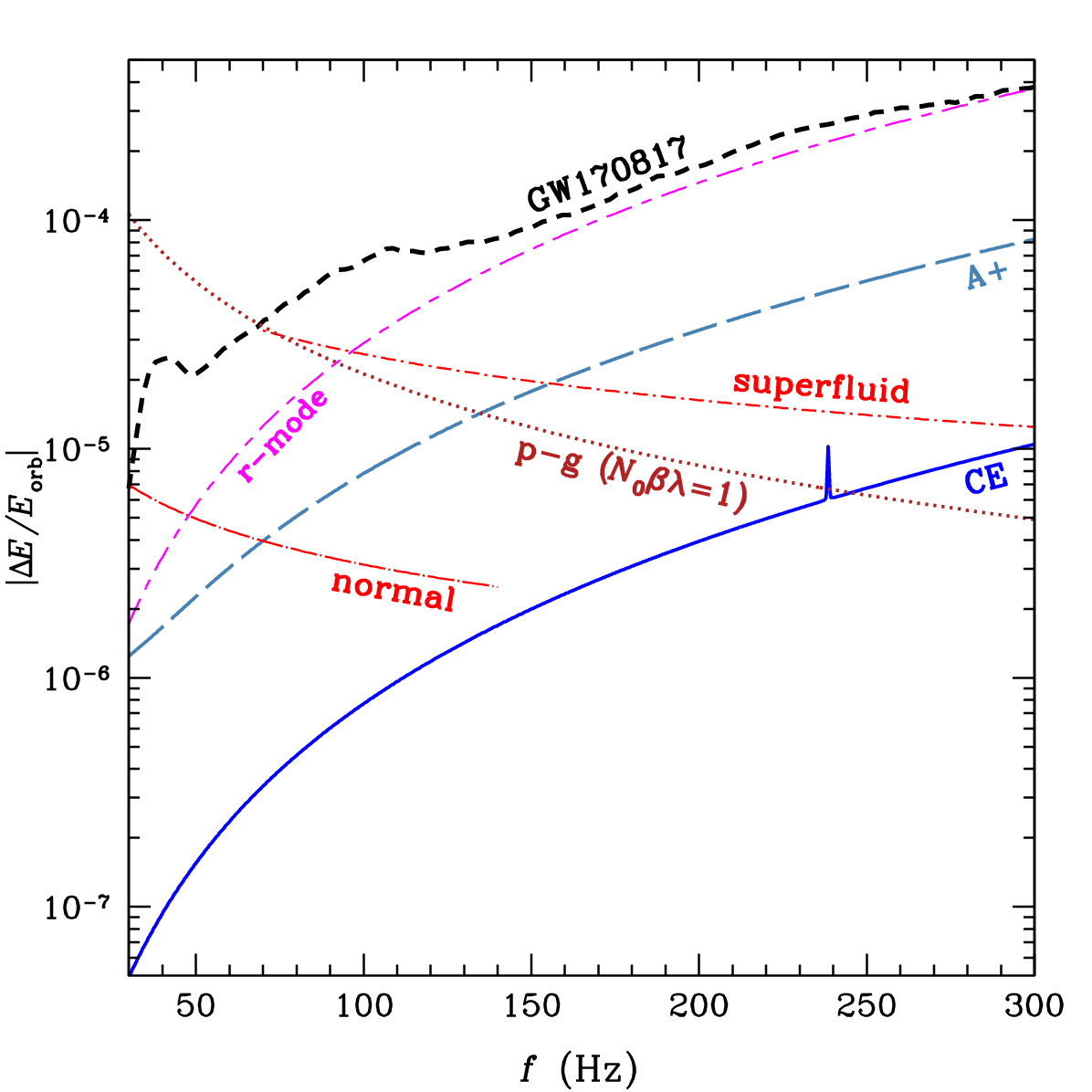}
\caption{
Relative energy transfer $|\Delta E/\Eorb|$ as a function of GW
frequency $\nugw$.
Thick lines are upper limits calculated using $\pm\Delphi$ from
Fig.~\ref{fig:dphi} and eq.~(\ref{eq:denergy})
for GW170817 (short-dashed), A+ (long-dashed), and Cosmic Explorer (CE; solid).
Long-dashed-dotted and short-dashed-dotted lines are for total
energy transferred to normal and superfluid g-modes, respectively
(see text for details),
while the dotted line is for total energy dissipated by the p-g
instability assuming NS radius $R=12\mbox{ km}$ and $\Npg\beta\lambda=1$
[see eq.~(\ref{eq:denergypg})]
and the short-long-dashed line is for energy transferred to the orbit
by r-modes and $R=12\mbox{ km}$ [see eq.~(\ref{eq:rmode})].
Note that for the r-mode, each particular GW frequency corresponds to
a specific NS spin frequency [here we assumed $\nugw=(4/3)\nus$].
\label{fig:denergy}}
\end{figure}

As noted in the Introduction, the phase uncertainties
shown in Figure~\ref{fig:dphi} are derived from comparisons between
the measured or expected GW data and waveform models.
Since these waveform models do not include dynamical tidal effects
such as the resonant excitation of oscillation modes at low frequencies
well before merger, $\Delphi$ from Figure~\ref{fig:dphi}
illustrates the potential to
constrain these unmodeled contributions to actual GW data.
In Figure~\ref{fig:denergy}, we show inferred upper limits on energy transfer
during the inspiral obtained by substituting
$\Delphi$ from Figure~\ref{fig:dphi} into equation~(\ref{eq:denergy}).
Also shown are the calculated total energy transferred to normal and
superfluid g-mode oscillations from \cite{yuweinberg17}
(for $l=2$ and summed over $n=1,2,\ldots,8$).
Note that the (model-dependent) frequencies of the $n=1$ and $n=8$
normal g-modes are approximately 140~Hz and 20~Hz, respectively, while the
corresponding frequencies for superfluid g-modes are 430~Hz and 70~Hz
\cite{yuweinberg17};
superfluid g-modes occur at higher frequencies than normal g-modes of
the same $n$ (by about a factor of the square root of proton fraction;
\cite{kantorgusakov14,passamontietal16}).

We see from Figure~\ref{fig:denergy} that the sensitivity of detectors
available at the time of GW170817 is insufficient
to observe any transfer of energy from the decaying orbit to either NS
with a limit of $\Delta E/|\Eorb|\approx10^{-5}-10^{-4}$ at
$\nugw>30\mbox{ Hz}$.
However, it seems possible that the sensitivity of A+ detectors
may be able to begin to measure energy transfers to normal
and superfluid g-modes in limited frequency ranges.
Meanwhile, CE could measure energy transfer to normal and superfluid g-modes
throughout the inspiral.

For resonance with a specific $m=2$ mode, the relation between
mode oscillation frequency $\omegamode$, orbital frequency $\Omegaorb$,
and gravitational wave frequency $\nugw$ is
\be
\omegamode = 2\Omegaorb = 2\pi\nugw.
\ee
For convenience, the mode oscillation frequency can be made dimensionless via
\be
\omegamode = \omegazero\omegamodend
 = 2\pi\times2170\mbox{ Hz }M_{1.4}^{1/2}R_{10}^{-3/2}\omegamodend,
 \label{eq:modefreq}
\ee
where $\omegazero=(GM_1/R^3)^{1/2}$ and $R_{10}=R/10\mbox{ km}$,
such that $\Omegaorb=\omegazero\omegamodend/2$ at resonance.
Note that we only consider here resonance with one of the NSs in a
binary NS system for simplicity.
The energy transfer due to resonant mode excitation is calculated
by \cite{lai94} to be
\ba
\frac{\Delta E}{|\Eorb|} &=& \frac{\pi^2}{128}\lp\frac{c^2R}{GM_1}\rp^{5/2}
 \lp\frac{2}{1+q}\rp^{4/3}\frac{\Qmode^2}{\omegamodend^{1/3}} \nonumber\\
&=& 11\,M_{1.4}^{-7/3}R_{10}^2\lp\frac{2}{1+q}\rp^{4/3}
 \Qmode^2\lp\frac{\nugw}{100\mbox{ Hz}}\rp^{-1/3}. \label{eq:denergymode}
\ea
Substituting equation~(\ref{eq:denergymode}) into equation~(\ref{eq:denergy})
gives
\ba
\frac{\Delphi}{2\pi}
&=& -\frac{5\pi}{4096}\lp\frac{c^2R}{GM_1}\rp^{5}\frac{2}{q(1+q)}
 \frac{\Qmode^2}{\omegamodend^2} \nonumber\\
&=& -4800\,M_{1.4}^{-4}R_{10}^2\frac{2}{q(1+q)}
 \Qmode^2\lp\frac{\nugw}{100\mbox{ Hz}}\rp^{-2}. \label{eq:qmode}
\ea
This phase shift occurs in a narrow frequency range since the duration
of the resonance is short compared to the inspiral duration in the
GW detector frequency band (see, e.g., \cite{lai94,flanaganracine07,poisson20}).

\begin{figure}
\includegraphics[width=0.45\textwidth]{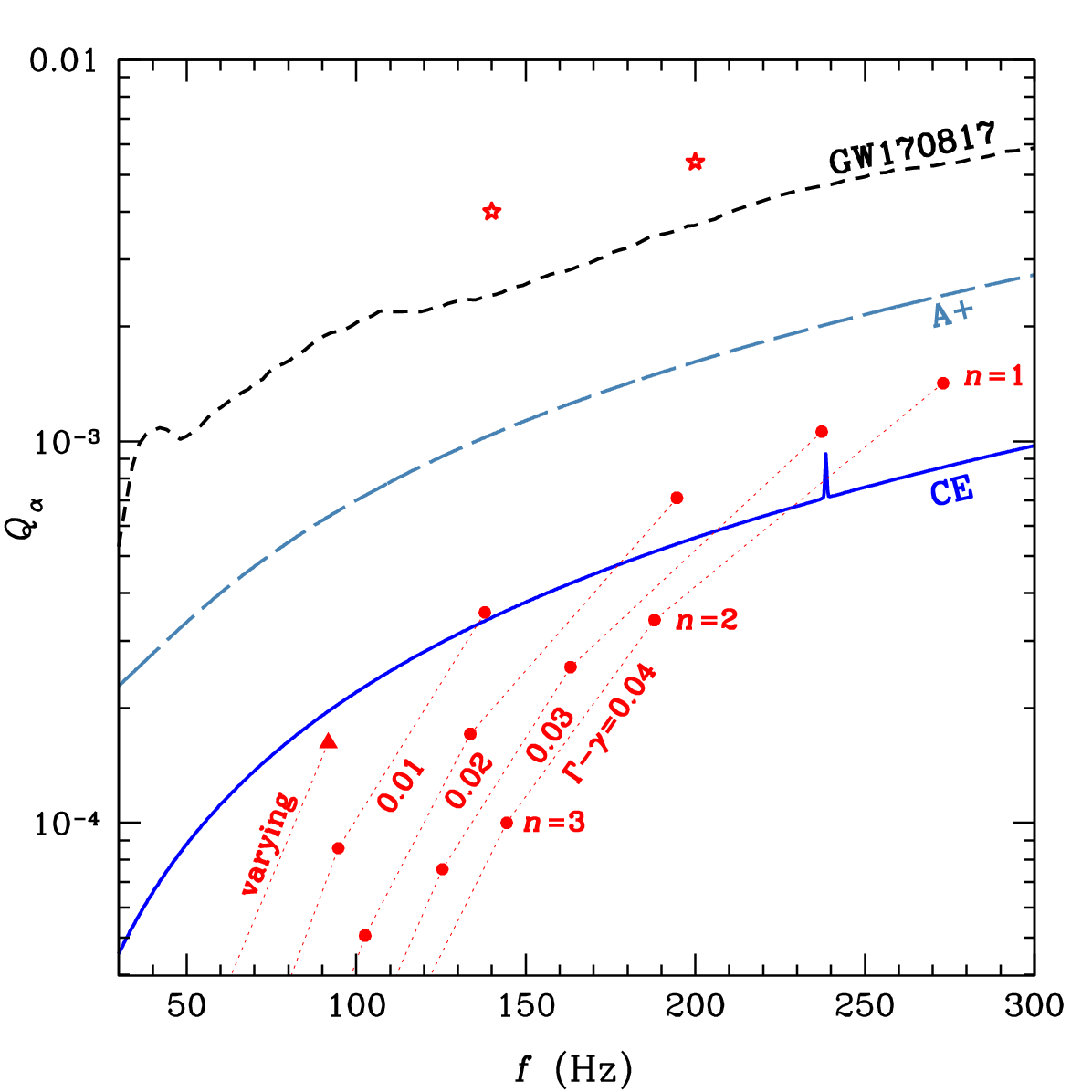}
\caption{
Dimensionless overlap integral $\Qmode$ as a function of GW frequency $\nugw$.
Lines are upper limits calculated using $\pm\Delphi$ from Fig.~\ref{fig:dphi}
and eq.~(\ref{eq:qmode}) with a NS radius $R=12\mbox{ km}$
for GW170817 (short-dashed), A+ (long-dashed), and Cosmic Explorer (CE; solid).
Circles are for constant $\Gamma-\gamma$ stratification g-modes,
while the triangle is for a varying stratification g-mode (see text);
light dotted lines connect $\Qmode$ values from g-modes with the same
stratification but different radial node $n$.
Stars are for g-modes ($n=1$) calculated from NS models motivated by
the BSk21 and SLy4 equations of state
(see Counsell et al., in prep., and \cite{yuweinberg17}, respectively),
which have strong internal composition gradients.
\label{fig:qmode}}
\end{figure}

Figure~\ref{fig:qmode} shows upper limits on the overlap integral
$\Qmode$ obtained by substituting $\Delphi$ from Figure~\ref{fig:dphi}
into equation~(\ref{eq:qmode}).
To illustrate how these limits compare approximately to theoretical
expectations, we plot values of $\Qmode$ obtained from NSs constructed
in Newtonian gravity
using a polytropic ($\gamma=2$) equation of state and g-modes produced
by density stratifications that are parameterized by the factor
$(\Gamma-\gamma)$, where $\Gamma$ is the adiabatic index;
results shown are for constant $(\Gamma-\gamma)$ (Counsell et al., in prep.),
which reproduce the results of \cite{xulai17},
and for a specific case of varying $(\Gamma-\gamma)$ from \cite{xulai17}.
\cite{xulai17} find the mode frequency and overlap integral scale as
$\omegamode \propto (M/R^3)^{1/2}(\Gamma-\gamma)^{1/2}$ and
$\Qmode \propto (\Gamma-\gamma)$ for a polytrope with a constant
$\gamma=2$, while
\cite{kuanetal21,kuanetal22} find an approximate scaling
$\omegamode \propto (M/R^4)^{1/3}(\Gamma-\gamma)^{1/2}$ for
NSs constructed using realistic equations of state but still with
g-mode stratifications parameterized by $(\Gamma-\gamma)$.
Also shown are sample results from Counsell et al. (in prep.) and
\cite{yuweinberg17} for $\Qmode$
determined using NS models which are motivated by the BSk21 and SLy4
equations of state and have strong varying stratifications.
In a sense, these two represent optimistic estimates of $\Qmode$, while
those from \cite{kuanetal22} with $\Qmode<10^{-4}$ are likely to be
conservative estimates, and the large range shows the level of uncertainty
in current theoretical calculations.
Meanwhile, we do not show $\Qmode$ for superfluid g-modes in
Figure~\ref{fig:qmode} since the ones from \cite{yuweinberg17} would
be below the CE curve.

While Figure~\ref{fig:denergy} suggests A+ and CE can
constrain or even measure the total energy transferred to many modes,
Figure~\ref{fig:qmode} shows that measuring the coupling to individual
modes via the overlap $\Qmode$ will be difficult even with CE,
although there are evident uncertainties in the theoretical calculations
of $\Qmode$, as discussed above.
Still, even in the somewhat pessimistic case, we may be able to draw
important conclusions about the NS interior.  For example,
non-detection of individual g-mode coupling with CE
could be used to constrain stratification within the NS,
e.g., $(\Gamma-\gamma)<0.02$,
or indicate neutrons in the NS core are superfluid.
Therefore works such as
\cite{yuweinberg17,rauwasserman18,kuanetal21,kuanetal22} are
in the right direction,
and more work is needed in using realistic equations of state and
stratification to calculate g-modes and their binary tidal interactions.
It is also important to keep in mind that $\Qmode$ from an extraordinary
event, such as an inspiralling NS closer than GW170817,
should be within the reach of CE.

\section{p-g instability\label{sec:pg}}

To quantify current and potential future constraints on the p-g
instability, we first estimate, from the rate of orbital energy
dissipation by the unstable modes $\dot{E}_{\rm pg}$
[see eq.~(3) of \cite{essicketal16}], that the total energy dissipated is
\ba
\frac{\Delta E}{|\Eorb|} &\sim& \frac{\dot{E}_{\rm pg}\tinspiral}{|\Eorb|}
= 10^{-8}\!\lp\frac{2}{1+q}\rp^{2/3}\!\!\lp\frac{2\pi\nugw}{\omegazero}\rp^{4/3}
 \!\!\!\omegazero\tinspiral \Npg\beta\lambda \nonumber\\
&=& 1.9\times10^{-5}\,M_{1.4}^{-11/6}R_{10}^{1/2}\frac{1}{q}
 \lp\frac{2}{1+q}\rp^{1/3} \nonumber\\
&& \times\lp\frac{\nugw}{100\mbox{ Hz}}\rp^{-4/3}\Npg\beta\lambda,
 \label{eq:denergypg}
\ea
where $\Npg$ is the number of independently unstable modes,
$\beta$ ($\le1$) indicates how close the energy at which unstable modes
saturate is to a maximum given by the wave breaking energy,
and $\lambda$ ($\sim0.1-1$) is a slowly varying function of binary separation
\cite{weinberg16}.
Figure~\ref{fig:denergy} plots equation~(\ref{eq:denergypg}) for
$\Npg\beta\lambda=1$,
which yields a limit $\Delta E<4\times10^{47}\mbox{ erg}$ at
$\nugw<70\mbox{ Hz}$ from GW170817.
It is argued in \cite{weinbergetal13,essicketal16,weinberg16}
that $\Npg\sim 100-10^4$, but we see that this is only possible if
$\beta\ll0.1$ and $\nugw\lesssim 70\mbox{ Hz}$ using just GW170817.
This limit would extend to all frequencies of importance for the
p-g instability ($\lesssim100\mbox{ Hz}$; \cite{essicketal16,abbottetal19})
by the time detectors reach A+ sensitivity.

\begin{figure*}[!t]
\includegraphics[width=0.45\textwidth]{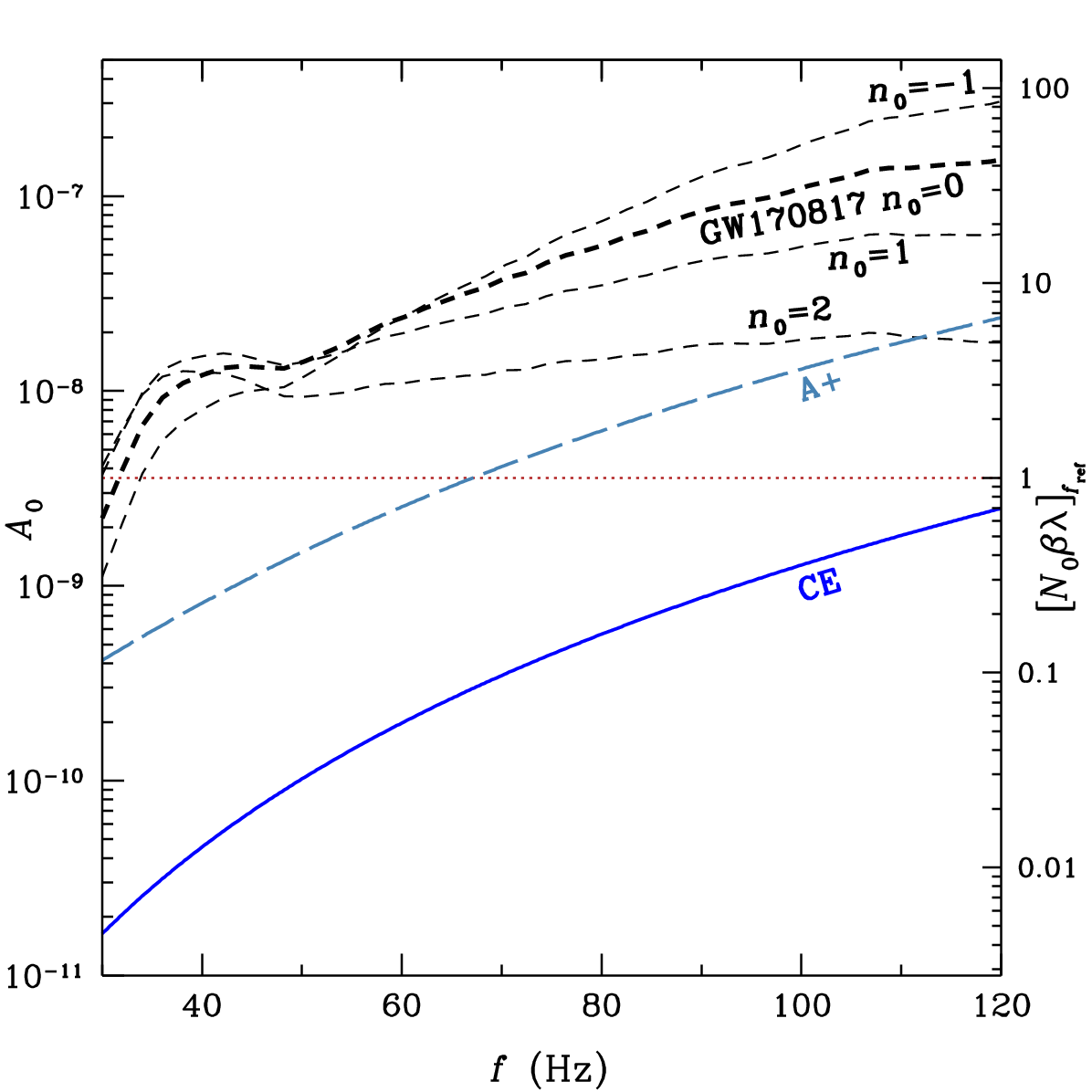}
\hspace{1cm}
\includegraphics[width=0.45\textwidth]{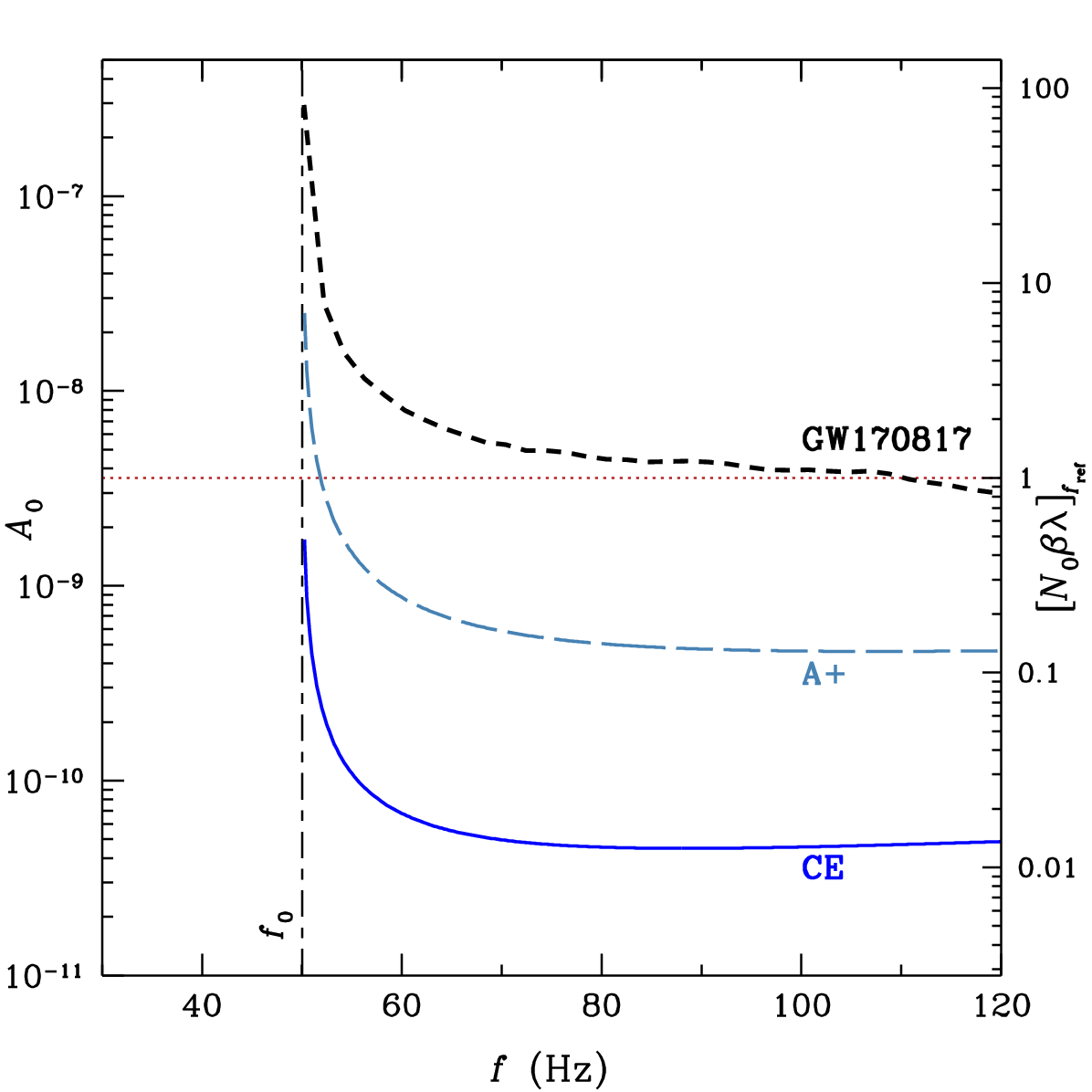}
\caption{
Amplitude of p-g instability $\Apg$ as a function of GW frequency $\nugw$.
Lines are upper limits calculated using $\pm\Delphi$ from Fig.~\ref{fig:dphi}
and the power law model given by eq.~(\ref{eq:pgdelphi}) (left panel)
and the asymptotic model given by eq.~(\ref{eq:pgdelphi16}) (right panel)
with frequency power law index $\npg=0$
for GW170817 (short-dashed; light lines are for $\npg=-1,+1,+2$),
A+ (long-dashed), and Cosmic Explorer (CE; solid).
The vertical line in the right panel indicates the frequency at which
mode saturation is assumed to occur, here taken to be $\nusat=50\mbox{ Hz}$.
Parameters of the p-g instability are the number of unstable modes $\Npg$
and mode energy relative to saturation maximum $\beta$ ($\le 1$)
and $\lambda$ ($\sim0.1-1$) is a function of binary separation.
\label{fig:pgdelphi}}
\end{figure*}

Next, we consider two parameterizations to determine the effect of
the p-g instability on the phase shift $\Delphi$.
The first from \cite{abbottetal19} is
\ba
\Delphi(\nugw>\nusat) &=&-\frac{2C_0}{3B^2(3-\npg)(4-\npg)}
 \lp\frac{\nugw}{\nuref}\rp^{\npg-3} \nonumber\\
&=& -2\pi\times 3.1\times10^6\,M_{1.4}^{-10/3}q^{-2} \nonumber\\
&& \times
 \frac{\Apg}{(3-\npg)(4-\npg)}\lp\frac{100\mbox{ Hz}}{\nugw}\rp^{3-\npg},
\label{eq:pgdelphi}
\ea
where $B=(32/5)(G\mchirp\pi\nuref/c^3)^{5/3}$, $C_0=[2/(1+q)]^{2/3}\Apg$,
$\nuref$ ($=100\mbox{ Hz}$ here) is an arbitrary reference frequency,
and $\npg$ describes the frequency scaling of the orbital energy
dissipation rate and is assumed to be in the range $-1\le\npg\le 3$.
Note that the full parameterization of \cite{abbottetal19} includes
dependences on the Heaviside function $\Theta(\nugw-\nusat)$,
where $\nusat$ ($\sim 50\mbox{ Hz}$; \cite{essicketal16,weinberg16})
is the mode saturation frequency,
but here we only kept the term that is non-zero at $\nugw>\nusat$.
In effect, we are ignoring the energy loss before the system reaches
saturation, which is reasonable provided the instability grows fast enough.
As in the resonance case,
we consider the phase contribution due to one NS rather than both.
One can relate the amplitude $\Apg$ to $\Npg\beta\lambda$ at $\nugw=\nuref$
as discussed in \cite{essicketal16}, i.e.,
\ba
\Apg &=& \lp\frac{2\pi\nuref}{\omegazero}\rp^{1/3}
 \lp\frac{\omega_g}{\Lambda_g\omegazero}\rp^2[\Npg\beta\lambda]_{\nuref}
 \nonumber\\
&=& 3.6\times10^{-9}\lp\frac{\omega_g}{10^{-4}\Lambda_g\omegazero}\rp^2
 [\Npg\beta\lambda]_{\nuref}, \label{eq:pgnbl}
\ea
where $\omega_g$ is the minimum g-mode frequency and $\Lambda_g=l(l+1)$.

The second parameterization is from \cite{essicketal16}:
\ba
\Delphi(\nugw>\nusat) &=&
 \frac{\Apg F_M}{\npg-3}\lp\frac{\nusat}{\nuref}\rp^{\npg-3}
 \lb1-\lp\frac{\nugw}{\nusat}\rp^{\npg-3}\rb \nonumber\\
&=& -2\pi\times 3.1\times10^6\,M_{1.4}^{-10/3}q^{-2} \nonumber\\
&& \times \frac{\Apg}{3-\npg} \lp\frac{100\mbox{ Hz}}{\nusat}\rp^{3-\npg}
 \lb1-\lp\frac{\nusat}{\nugw}\rp^{3-\npg}\rb, \label{eq:pgdelphi16}
\ea
where $\Apg F_M=2C_0/3B^2$.
Although \cite{abbottetal19} points out that this parameterization
was calculated incorrectly, it is useful to compare the two since
they can be interpreted more generally as models for unknown processes
that could affect GW signals, in particular, the former is simply a
power law and the latter is asymptotic to a constant phase shift.
Specific to the p-g instability, the above two parameterizations
of $\Delphi$ differ in amplitude by $\sim(4-\npg)(\nugw/\nusat)^{3-\npg}$
[cf. $(4-\npg)$ as stated in \cite{abbottetal19}].
In the more general case, they have very different frequency behavior.
Equation~(\ref{eq:pgdelphi}) implies that $\Delphi\propto (1/f)^{3-\npg}$,
so that $\Apg\propto f^{3-\npg}$ for a given $\Delphi$,
and hence $\Delphi$ ($\Apg$) decreases (increases) as the frequency
increases for $\npg<3$.
On the other hand, equation~(\ref{eq:pgdelphi16}) states that
$\Delphi\propto[1-(\nusat/f)^{3-\npg}]$, so that
$\Delphi$ ($\Apg$) increases (decreases) as the frequency increases
for $\npg<3$ and asymptotes to a constant when $\nugw\gg\nusat$.

The left and right panels of
Figure~\ref{fig:pgdelphi} show the upper limits on the amplitude
$\Apg$ (and $\Npg\beta\lambda$) obtained by substituting $\Delphi$
from Figure~\ref{fig:dphi} into equations~(\ref{eq:pgdelphi})
and (\ref{eq:pgdelphi16}), respectively.
One clearly sees the significantly different amplitudes and
frequency behavior described above.
Note that the small changes in $\Apg$ at $\nugw\gg\nusat$ in the
right panel are due to the frequency dependence of the measured/expected
$\Delphi$ from Figure~\ref{fig:dphi}.

Focusing only on the power law results from the more recent, corrected
parameterization of the p-g instability shown in the left panel, we see
that the amplitude $\Apg$ is already constrained using GW170817 data
to $\Apg<10^{-7}$ for $\npg>-1$ at frequencies most relevant to the
instability, i.e., $\nugw\lesssim 100\mbox{ Hz}$.
Converting this amplitude $\Apg$ to $\Npg\beta\lambda$ using
equation~(\ref{eq:pgnbl}) and assuming $\beta\sim1$ and $\lambda\sim1$,
the number of unstable p-g modes is limited to $\Npg\ll 100$ just from
GW170817 and could be as low as $\Npg\sim\mbox{a few}$ at
$\nugw\lesssim50\mbox{ Hz}$ and/or for $\npg\gtrsim2$.
The constraint from the amplitude $\Apg$ here is weaker than from
energy loss $\Delta E$ shown in Figure~\ref{fig:denergy},
where the latter gives $\Npg\beta\lambda<1$ at $\nugw<70\mbox{ Hz}$
using GW170817.
While there are many uncertainties with the physics
of the p-g instability and with our estimates, the p-g instability
appears to be much less effective than first suggested.  In fact, it may be
inconsequential in the evolution of inspiralling NSs.
Our analysis can also be interpreted as constraining the amplitude of
undetermined mechanisms that are described by the two parameterization
models considered above.

A detailed study of GW data from GW170817 specifically searching 
for effects of the p-g instability was conducted in \cite{abbottetal19}.
The analysis did not provide evidence for effects of the instability
but, if assumed to be present, obtained constraints on its parameters
$\nusat$, $\npg$, and $\Apg$.
The search for the mode saturation frequency $\nusat$ in the range
10--100~Hz yielded a peak in the posterior distribution at
$\sim 70\mbox{ Hz}$, but such a peak also appeared in simulated data and
thus was attributed to noise.
The results indicated a slight preference for $\npg>2$.
An upper limit of $\Apg\approx(3-7)\times10^{-7}$ was obtained, and
hence $\Npg\lesssim 200$ at 100~Hz, assuming $\beta=1$ and $\lambda=1$.
The energy dissipated by the p-g instability was constrained to be
$\Delta E<2\times 10^{48}\mbox{ erg}$ (90\% confidence)
at $\nugw\le70\mbox{ Hz}$.
These constraints on $\Apg$, $\Npg$, and $\Delta E$ from the more
comprehensive study are all within a factor of several of our simple
estimates.

\section{Discussion\label{sec:discuss}}

In this work, we showed that recently calculated uncertainties of the
phase of GW signals detected from the binary NS inspiral and merger
GW170817 and similar uncertainties expected from the future detectors of A+
and CE (and presumably ET) are much smaller than prior work assumed
at GW frequencies $\nugw\le 300\mbox{ Hz}$, where differences in
theoretical waveforms are not significant.
These smaller $\Delphi$ suggest future detections of
nearby merging NSs could begin to measure additional orbital energy
loss via mechanisms such as excitation of NS g-mode oscillations.
However, the precision in $\Delphi$ may be insufficient to
enable measurements of the strength of the coupling between the tidal
potential and individual oscillation modes $\Qmode$, unless a very
nearby NS inspiral is detected.
But it is important to be mindful of the fact that our current values of
$\Qmode$ are uncertain due to theoretical uncertainties and approximations.
In order to make progress, we need accurate relativistic calculations
for realistic NS models.
At the least, non-detections with CE may provide constraints on
particle fractions or superfluid matter inside NSs.
We also find that the phase uncertainty of the signal from GW170817
compared to the phase changes implied by the dissipated energy and
amplitude of the p-g instability already limits the number of unstable
p-g modes to as low as $\Npg\sim 1$ if the other parameters are at values
expected from the model,
i.e., $\beta\sim1$ and $\lambda\approx0.1-1$.
Such low values of $\Npg$ indicate the p-g instability is unlikely to
produce interesting effects in GW signals from NS mergers, although
there are many unknowns here as well.

One avenue that might lead to improved sensitivity to detecting dynamical
tidal effects like those studied here is to combine GW signals from
multiple inspiralling NS systems.
Indeed there are studies of stacking signals to determine how well
masses and tidal deformabilities could be measured
(see, e.g., \cite{agathosetal15}).
The strategy for the case here would be more complicated as it also
involves matter composition, but it is clearly worth exploring.
To measure resonances at specific GW frequencies from a stacked signal
would require that, e.g., the relation between g-mode frequencies and
mass, radius, and stratification are known precisely.

We focused on effects at GW frequencies below 300~Hz because
comparisons between current waveforms used to analyze GW data of
merging NSs have differences that exceed the measured or expected
$\Delphi$ at higher frequencies \cite{read23}.
This aspect also needs to be addressed.
As waveform systematic uncertainties are improved in the coming years,
$\Delphi$ at higher frequencies would become limited only by physics
not included in the waveforms.  Thus these more precise waveforms
could be used to constrain the presence of unmodeled physics like
those studied here but at higher frequencies, e.g., g-modes in superfluid NSs.

Finally,
in Section~\ref{sec:gmode}, we considered orbital energy loss and GW
phase shifts due to resonant excitation of g-mode oscillations during
the inspiral.
Another potential fluid oscillation that could be resonantly excited
gravitomagnetically is a r-mode oscillation
\cite{anderssonkokkotas01,flanaganracine07}.
The phase shift from r-mode resonance is estimated by
\cite{flanaganracine07} (see also \cite{poisson20}) to be
\be
\frac{\Delphi}{2\pi} = 0.006\,M_{1.4}^{-10/3}R_{10}^4\frac{1}{q}
 \!\lp\frac{2}{1+q}\rp^{1/3}\!\!\lp\frac{\nugw}{100\mbox{ Hz}}\rp^{2/3}.
 \label{eq:rmode}
\ee
Figure~\ref{fig:denergy} shows the result of substituting this into
equation~(\ref{eq:denergy}).
We see that the effect of r-mode resonances may be quite significant
and could be detectable in the near future.
This motivated recent, more detailed calculations, performed in order to
estimate how well NS parameters could be extracted from data
\cite{maetal21,guptaetal23}.

There are several important points to note regarding tidal excitation
of r-modes.
First, the above r-mode phase shift implies energy is transferred to the
orbit such that orbital decay is prolonged, although \cite{poisson20}
finds there are several related mode resonances that could lead to phase
shifts in the opposite direction, depending on the alignment between
orbital and NS spin angular momenta.
Second, the frequency
at which resonance, and thus the phase shift, occurs is proportional
to the NS spin rate $\nus$, i.e., $\nugw=(4/3)\nus$ with a slightly
larger coefficient after accounting for relativistic effects that
depend on $M/R$ \cite{lockitchetal03,idrisyetal15}.
Therefore the NS must be rapidly rotating in order for the
resonance to come into effect in the sensitivity band of ground-based detectors.
Based on our current understanding and observations of Galactic systems,
NSs in binaries that will merge in a Hubble time have spin rates
$\nus<50\mbox{ Hz}$.
For an inspiralling NS-NS system, it is probably unlikely that both NSs
would have high enough spin rates for each to have their r-mode resonantly
excited during the late stages of inspiral.
Most importantly though, detection of a phase delay at a specific
frequency would provide independent determinations of $M$, $R$, and $\nus$,
which would complement parameters extracted from the conventional
inspiral signal and thus could be used to break degeneracies.
Even non-detection could be indicative of a merging NS (or NSs)
that has a low spin rate.
Thus, while more challenging at lower GW frequencies, detection is
possible and an exciting prospect for the era of A+, CE, and ET.

\begin{acknowledgments}
The authors are grateful to Bruce Edelman, Ben Farr, and Jocelyn Read
for providing the phase uncertainties from their works
(and shown in Figure~\ref{fig:dphi}) and to Jocelyn Read for her work
and discussions.
The authors thank Andrew Counsell for sharing his numerical g-mode
results with us.
NA gratefully acknowledges support from Science and Technology Facility
Council (STFC) via grant number ST/V000551/1.
\end{acknowledgments}

\bibliography{tides23}

\begin{thebibliography}{31}%
\makeatletter
\providecommand \@ifxundefined [1]{%
 \@ifx{#1\undefined}
}%
\providecommand \@ifnum [1]{%
 \ifnum #1\expandafter \@firstoftwo
 \else \expandafter \@secondoftwo
 \fi
}%
\providecommand \@ifx [1]{%
 \ifx #1\expandafter \@firstoftwo
 \else \expandafter \@secondoftwo
 \fi
}%
\providecommand \natexlab [1]{#1}%
\providecommand \enquote  [1]{``#1''}%
\providecommand \bibnamefont  [1]{#1}%
\providecommand \bibfnamefont [1]{#1}%
\providecommand \citenamefont [1]{#1}%
\providecommand \href@noop [0]{\@secondoftwo}%
\providecommand \href [0]{\begingroup \@sanitize@url \@href}%
\providecommand \@href[1]{\@@startlink{#1}\@@href}%
\providecommand \@@href[1]{\endgroup#1\@@endlink}%
\providecommand \@sanitize@url [0]{\catcode `\\12\catcode `\$12\catcode
  `\&12\catcode `\#12\catcode `\^12\catcode `\_12\catcode `\%12\relax}%
\providecommand \@@startlink[1]{}%
\providecommand \@@endlink[0]{}%
\providecommand \url  [0]{\begingroup\@sanitize@url \@url }%
\providecommand \@url [1]{\endgroup\@href {#1}{\urlprefix }}%
\providecommand \urlprefix  [0]{URL }%
\providecommand \Eprint [0]{\href }%
\providecommand \doibase [0]{https://doi.org/}%
\providecommand \selectlanguage [0]{\@gobble}%
\providecommand \bibinfo  [0]{\@secondoftwo}%
\providecommand \bibfield  [0]{\@secondoftwo}%
\providecommand \translation [1]{[#1]}%
\providecommand \BibitemOpen [0]{}%
\providecommand \bibitemStop [0]{}%
\providecommand \bibitemNoStop [0]{.\EOS\space}%
\providecommand \EOS [0]{\spacefactor3000\relax}%
\providecommand \BibitemShut  [1]{\csname bibitem#1\endcsname}%
\let\auto@bib@innerbib\@empty
\bibitem [{\citenamefont {{B.~P. {Abbott} {\it et al.}}}(2017)}]{abbottetal17}%
  \BibitemOpen
  \bibfield  {author} {\bibinfo {author} {\bibnamefont {{B.~P. {Abbott} {\it et
  al.}}}},\ }\href {https://doi.org/10.1103/PhysRevLett.119.161101} {\bibfield
  {journal} {\bibinfo  {journal} {\prl}\ }\textbf {\bibinfo {volume} {119}},\
  \bibinfo {eid} {161101} (\bibinfo {year} {2017})}\BibitemShut {NoStop}%
\bibitem [{\citenamefont {{B.~P. {Abbott} {\it et al.}}}(2018)}]{abbottetal18}%
  \BibitemOpen
  \bibfield  {author} {\bibinfo {author} {\bibnamefont {{B.~P. {Abbott} {\it et
  al.}}}},\ }\href {https://doi.org/10.1103/PhysRevLett.121.161101} {\bibfield
  {journal} {\bibinfo  {journal} {\prl}\ }\textbf {\bibinfo {volume} {121}},\
  \bibinfo {eid} {161101} (\bibinfo {year} {2018})}\BibitemShut {NoStop}%
\bibitem [{\citenamefont {{Read}}(2023)}]{read23}%
  \BibitemOpen
  \bibfield  {author} {\bibinfo {author} {\bibfnamefont {J.}~\bibnamefont
  {{Read}}},\ }\href {https://doi.org/10.1088/1361-6382/acd29b} {\bibfield
  {journal} {\bibinfo  {journal} {Classical and Quantum Gravity}\ }\textbf
  {\bibinfo {volume} {40}},\ \bibinfo {eid} {135002} (\bibinfo {year}
  {2023})}\BibitemShut {NoStop}%
\bibitem [{\citenamefont {{Kantor}}\ and\ \citenamefont
  {{Gusakov}}(2014)}]{kantorgusakov14}%
  \BibitemOpen
  \bibfield  {author} {\bibinfo {author} {\bibfnamefont {E.~M.}\ \bibnamefont
  {{Kantor}}}\ and\ \bibinfo {author} {\bibfnamefont {M.~E.}\ \bibnamefont
  {{Gusakov}}},\ }\href {https://doi.org/10.1093/mnrasl/slu061} {\bibfield
  {journal} {\bibinfo  {journal} {\mnras}\ }\textbf {\bibinfo {volume} {442}},\
  \bibinfo {pages} {L90} (\bibinfo {year} {2014})}\BibitemShut {NoStop}%
\bibitem [{\citenamefont {{Andersson}}\ and\ \citenamefont
  {{Ho}}(2018)}]{anderssonho18}%
  \BibitemOpen
  \bibfield  {author} {\bibinfo {author} {\bibfnamefont {N.}~\bibnamefont
  {{Andersson}}}\ and\ \bibinfo {author} {\bibfnamefont {W.~C.~G.}\
  \bibnamefont {{Ho}}},\ }\href {https://doi.org/10.1103/PhysRevD.97.023016}
  {\bibfield  {journal} {\bibinfo  {journal} {\prd}\ }\textbf {\bibinfo
  {volume} {97}},\ \bibinfo {eid} {023016} (\bibinfo {year}
  {2018})}\BibitemShut {NoStop}%
\bibitem [{\citenamefont {{Press}}\ and\ \citenamefont
  {{Teukolsky}}(1977)}]{pressteukolsky77}%
  \BibitemOpen
  \bibfield  {author} {\bibinfo {author} {\bibfnamefont {W.~H.}\ \bibnamefont
  {{Press}}}\ and\ \bibinfo {author} {\bibfnamefont {S.~A.}\ \bibnamefont
  {{Teukolsky}}},\ }\href {https://doi.org/10.1086/155143} {\bibfield
  {journal} {\bibinfo  {journal} {\apj}\ }\textbf {\bibinfo {volume} {213}},\
  \bibinfo {pages} {183} (\bibinfo {year} {1977})}\BibitemShut {NoStop}%
\bibitem [{\citenamefont {{Lai}}(1994)}]{lai94}%
  \BibitemOpen
  \bibfield  {author} {\bibinfo {author} {\bibfnamefont {D.}~\bibnamefont
  {{Lai}}},\ }\href
  {https://doi.org/10.1093/mnras/270.3.61110.48550/arXiv.astro-ph/9404062}
  {\bibfield  {journal} {\bibinfo  {journal} {\mnras}\ }\textbf {\bibinfo
  {volume} {270}},\ \bibinfo {pages} {611} (\bibinfo {year}
  {1994})}\BibitemShut {NoStop}%
\bibitem [{\citenamefont {{Kokkotas}}\ and\ \citenamefont
  {{Schafer}}(1995)}]{kokkotasschafer95}%
  \BibitemOpen
  \bibfield  {author} {\bibinfo {author} {\bibfnamefont {K.~D.}\ \bibnamefont
  {{Kokkotas}}}\ and\ \bibinfo {author} {\bibfnamefont {G.}~\bibnamefont
  {{Schafer}}},\ }\href {https://doi.org/10.1093/mnras/275.2.301} {\bibfield
  {journal} {\bibinfo  {journal} {\mnras}\ }\textbf {\bibinfo {volume} {275}},\
  \bibinfo {pages} {301} (\bibinfo {year} {1995})}\BibitemShut {NoStop}%
\bibitem [{\citenamefont {{Reisenegger}}\ and\ \citenamefont
  {{Goldreich}}(1994)}]{reiseneggergoldreich94}%
  \BibitemOpen
  \bibfield  {author} {\bibinfo {author} {\bibfnamefont {A.}~\bibnamefont
  {{Reisenegger}}}\ and\ \bibinfo {author} {\bibfnamefont {P.}~\bibnamefont
  {{Goldreich}}},\ }\href {https://doi.org/10.1086/174105} {\bibfield
  {journal} {\bibinfo  {journal} {\apj}\ }\textbf {\bibinfo {volume} {426}},\
  \bibinfo {pages} {688} (\bibinfo {year} {1994})}\BibitemShut {NoStop}%
\bibitem [{\citenamefont {{Weinberg}}\ \emph {et~al.}(2013)\citenamefont
  {{Weinberg}}, \citenamefont {{Arras}},\ and\ \citenamefont
  {{Burkart}}}]{weinbergetal13}%
  \BibitemOpen
  \bibfield  {author} {\bibinfo {author} {\bibfnamefont {N.~N.}\ \bibnamefont
  {{Weinberg}}}, \bibinfo {author} {\bibfnamefont {P.}~\bibnamefont
  {{Arras}}},\ and\ \bibinfo {author} {\bibfnamefont {J.}~\bibnamefont
  {{Burkart}}},\ }\href {https://doi.org/10.1088/0004-637X/769/2/121}
  {\bibfield  {journal} {\bibinfo  {journal} {\apj}\ }\textbf {\bibinfo
  {volume} {769}},\ \bibinfo {eid} {121} (\bibinfo {year} {2013})}\BibitemShut
  {NoStop}%
\bibitem [{\citenamefont {{Weinberg}}(2016)}]{weinberg16}%
  \BibitemOpen
  \bibfield  {author} {\bibinfo {author} {\bibfnamefont {N.~N.}\ \bibnamefont
  {{Weinberg}}},\ }\href
  {https://doi.org/10.3847/0004-637X/819/2/10910.48550/arXiv.1509.06975}
  {\bibfield  {journal} {\bibinfo  {journal} {\apj}\ }\textbf {\bibinfo
  {volume} {819}},\ \bibinfo {eid} {109} (\bibinfo {year} {2016})}\BibitemShut
  {NoStop}%
\bibitem [{\citenamefont {{Essick}}\ \emph {et~al.}(2016)\citenamefont
  {{Essick}}, \citenamefont {{Vitale}},\ and\ \citenamefont
  {{Weinberg}}}]{essicketal16}%
  \BibitemOpen
  \bibfield  {author} {\bibinfo {author} {\bibfnamefont {R.}~\bibnamefont
  {{Essick}}}, \bibinfo {author} {\bibfnamefont {S.}~\bibnamefont {{Vitale}}},\
  and\ \bibinfo {author} {\bibfnamefont {N.~N.}\ \bibnamefont {{Weinberg}}},\
  }\href {https://doi.org/10.1103/PhysRevD.94.10301210.48550/arXiv.1609.06362}
  {\bibfield  {journal} {\bibinfo  {journal} {\prd}\ }\textbf {\bibinfo
  {volume} {94}},\ \bibinfo {eid} {103012} (\bibinfo {year}
  {2016})}\BibitemShut {NoStop}%
\bibitem [{\citenamefont {{B.~P. {Abbott} {\it et al.}}}(2019)}]{abbottetal19}%
  \BibitemOpen
  \bibfield  {author} {\bibinfo {author} {\bibnamefont {{B.~P. {Abbott} {\it et
  al.}}}},\ }\href
  {https://doi.org/10.1103/PhysRevLett.122.06110410.48550/arXiv.1808.08676}
  {\bibfield  {journal} {\bibinfo  {journal} {\prl}\ }\textbf {\bibinfo
  {volume} {122}},\ \bibinfo {eid} {061104} (\bibinfo {year}
  {2019})}\BibitemShut {NoStop}%
\bibitem [{\citenamefont {{Cutler}}\ and\ \citenamefont
  {{Flanagan}}(1994)}]{cutlerflanagan94}%
  \BibitemOpen
  \bibfield  {author} {\bibinfo {author} {\bibfnamefont {C.}~\bibnamefont
  {{Cutler}}}\ and\ \bibinfo {author} {\bibfnamefont {{\'E}.~E.}\ \bibnamefont
  {{Flanagan}}},\ }\href {https://doi.org/10.1103/PhysRevD.49.2658} {\bibfield
  {journal} {\bibinfo  {journal} {\prd}\ }\textbf {\bibinfo {volume} {49}},\
  \bibinfo {pages} {2658} (\bibinfo {year} {1994})}\BibitemShut {NoStop}%
\bibitem [{\citenamefont {{Balachandran}}\ and\ \citenamefont
  {{Flanagan}}(2007)}]{balachandranflanagan07}%
  \BibitemOpen
  \bibfield  {author} {\bibinfo {author} {\bibfnamefont {P.}~\bibnamefont
  {{Balachandran}}}\ and\ \bibinfo {author} {\bibfnamefont {E.~E.}\
  \bibnamefont {{Flanagan}}},\ }\href
  {https://doi.org/10.48550/arXiv.gr-qc/0701076} {\bibfield  {journal}
  {\bibinfo  {journal} {arXiv e-prints}\ ,\ \bibinfo {eid} {gr-qc/0701076}}
  (\bibinfo {year} {2007})}\BibitemShut {NoStop}%
\bibitem [{\citenamefont {{Edelman}}\ \emph {et~al.}(2021)\citenamefont
  {{Edelman}}, \citenamefont {{Rivera-Paleo}}, \citenamefont {{Merritt}},
  \citenamefont {{Farr}}, \citenamefont {{Doctor}}, \citenamefont {{Brink}},
  \citenamefont {{Farr}}, \citenamefont {{Gair}}, \citenamefont {{Key}},
  \citenamefont {{McIver}},\ and\ \citenamefont {{Nielsen}}}]{edelmanetal21}%
  \BibitemOpen
  \bibfield  {author} {\bibinfo {author} {\bibfnamefont {B.}~\bibnamefont
  {{Edelman}}}, \bibinfo {author} {\bibfnamefont {F.~J.}\ \bibnamefont
  {{Rivera-Paleo}}}, \bibinfo {author} {\bibfnamefont {J.~D.}\ \bibnamefont
  {{Merritt}}}, \bibinfo {author} {\bibfnamefont {B.}~\bibnamefont {{Farr}}},
  \bibinfo {author} {\bibfnamefont {Z.}~\bibnamefont {{Doctor}}}, \bibinfo
  {author} {\bibfnamefont {J.}~\bibnamefont {{Brink}}}, \bibinfo {author}
  {\bibfnamefont {W.~M.}\ \bibnamefont {{Farr}}}, \bibinfo {author}
  {\bibfnamefont {J.}~\bibnamefont {{Gair}}}, \bibinfo {author} {\bibfnamefont
  {J.~S.}\ \bibnamefont {{Key}}}, \bibinfo {author} {\bibfnamefont
  {J.}~\bibnamefont {{McIver}}},\ and\ \bibinfo {author} {\bibfnamefont
  {A.~B.}\ \bibnamefont {{Nielsen}}},\ }\href
  {https://doi.org/10.1103/PhysRevD.103.04200410.48550/arXiv.2008.06436}
  {\bibfield  {journal} {\bibinfo  {journal} {\prd}\ }\textbf {\bibinfo
  {volume} {103}},\ \bibinfo {eid} {042004} (\bibinfo {year}
  {2021})}\BibitemShut {NoStop}%
\bibitem [{\citenamefont {{Dietrich}}\ \emph {et~al.}(2019)\citenamefont
  {{Dietrich}}, \citenamefont {{Khan}}, \citenamefont {{Dudi}}, \citenamefont
  {{Kapadia}}, \citenamefont {{Kumar}}, \citenamefont {{Nagar}}, \citenamefont
  {{Ohme}}, \citenamefont {{Pannarale}}, \citenamefont {{Samajdar}},
  \citenamefont {{Bernuzzi}}, \citenamefont {{Carullo}}, \citenamefont {{Del
  Pozzo}}, \citenamefont {{Haney}}, \citenamefont {{Markakis}}, \citenamefont
  {{P{\"u}rrer}}, \citenamefont {{Riemenschneider}}, \citenamefont
  {{Setyawati}}, \citenamefont {{Tsang}},\ and\ \citenamefont {{Van Den
  Broeck}}}]{dietrichetal19}%
  \BibitemOpen
  \bibfield  {author} {\bibinfo {author} {\bibfnamefont {T.}~\bibnamefont
  {{Dietrich}}}, \bibinfo {author} {\bibfnamefont {S.}~\bibnamefont {{Khan}}},
  \bibinfo {author} {\bibfnamefont {R.}~\bibnamefont {{Dudi}}}, \bibinfo
  {author} {\bibfnamefont {S.~J.}\ \bibnamefont {{Kapadia}}}, \bibinfo {author}
  {\bibfnamefont {P.}~\bibnamefont {{Kumar}}}, \bibinfo {author} {\bibfnamefont
  {A.}~\bibnamefont {{Nagar}}}, \bibinfo {author} {\bibfnamefont
  {F.}~\bibnamefont {{Ohme}}}, \bibinfo {author} {\bibfnamefont
  {F.}~\bibnamefont {{Pannarale}}}, \bibinfo {author} {\bibfnamefont
  {A.}~\bibnamefont {{Samajdar}}}, \bibinfo {author} {\bibfnamefont
  {S.}~\bibnamefont {{Bernuzzi}}}, \bibinfo {author} {\bibfnamefont
  {G.}~\bibnamefont {{Carullo}}}, \bibinfo {author} {\bibfnamefont
  {W.}~\bibnamefont {{Del Pozzo}}}, \bibinfo {author} {\bibfnamefont
  {M.}~\bibnamefont {{Haney}}}, \bibinfo {author} {\bibfnamefont
  {C.}~\bibnamefont {{Markakis}}}, \bibinfo {author} {\bibfnamefont
  {M.}~\bibnamefont {{P{\"u}rrer}}}, \bibinfo {author} {\bibfnamefont
  {G.}~\bibnamefont {{Riemenschneider}}}, \bibinfo {author} {\bibfnamefont
  {Y.~E.}\ \bibnamefont {{Setyawati}}}, \bibinfo {author} {\bibfnamefont
  {K.~W.}\ \bibnamefont {{Tsang}}},\ and\ \bibinfo {author} {\bibfnamefont
  {C.}~\bibnamefont {{Van Den Broeck}}},\ }\href
  {https://doi.org/10.1103/PhysRevD.99.024029} {\bibfield  {journal} {\bibinfo
  {journal} {\prd}\ }\textbf {\bibinfo {volume} {99}},\ \bibinfo {eid} {024029}
  (\bibinfo {year} {2019})}\BibitemShut {NoStop}%
\bibitem [{\citenamefont {{Ma}}\ \emph {et~al.}(2021)\citenamefont {{Ma}},
  \citenamefont {{Yu}},\ and\ \citenamefont {{Chen}}}]{maetal21}%
  \BibitemOpen
  \bibfield  {author} {\bibinfo {author} {\bibfnamefont {S.}~\bibnamefont
  {{Ma}}}, \bibinfo {author} {\bibfnamefont {H.}~\bibnamefont {{Yu}}},\ and\
  \bibinfo {author} {\bibfnamefont {Y.}~\bibnamefont {{Chen}}},\ }\href
  {https://doi.org/10.1103/PhysRevD.103.063020} {\bibfield  {journal} {\bibinfo
   {journal} {\prd}\ }\textbf {\bibinfo {volume} {103}},\ \bibinfo {eid}
  {063020} (\bibinfo {year} {2021})}\BibitemShut {NoStop}%
\bibitem [{\citenamefont {{Gupta}}\ \emph {et~al.}(2023)\citenamefont
  {{Gupta}}, \citenamefont {{Steinhoff}},\ and\ \citenamefont
  {{Hinderer}}}]{guptaetal23}%
  \BibitemOpen
  \bibfield  {author} {\bibinfo {author} {\bibfnamefont {P.~K.}\ \bibnamefont
  {{Gupta}}}, \bibinfo {author} {\bibfnamefont {J.}~\bibnamefont
  {{Steinhoff}}},\ and\ \bibinfo {author} {\bibfnamefont {T.}~\bibnamefont
  {{Hinderer}}},\ }\href {https://doi.org/10.48550/arXiv.2302.11274} {\bibfield
   {journal} {\bibinfo  {journal} {arXiv e-prints}\ ,\ \bibinfo {eid}
  {arXiv:2302.11274}} (\bibinfo {year} {2023})}\BibitemShut {NoStop}%
\bibitem [{\citenamefont {{Yu}}\ and\ \citenamefont
  {{Weinberg}}(2017)}]{yuweinberg17}%
  \BibitemOpen
  \bibfield  {author} {\bibinfo {author} {\bibfnamefont {H.}~\bibnamefont
  {{Yu}}}\ and\ \bibinfo {author} {\bibfnamefont {N.~N.}\ \bibnamefont
  {{Weinberg}}},\ }\href
  {https://doi.org/10.1093/mnras/stw255210.48550/arXiv.1610.00745} {\bibfield
  {journal} {\bibinfo  {journal} {\mnras}\ }\textbf {\bibinfo {volume} {464}},\
  \bibinfo {pages} {2622} (\bibinfo {year} {2017})}\BibitemShut {NoStop}%
\bibitem [{\citenamefont {{Passamonti}}\ \emph {et~al.}(2016)\citenamefont
  {{Passamonti}}, \citenamefont {{Andersson}},\ and\ \citenamefont
  {{Ho}}}]{passamontietal16}%
  \BibitemOpen
  \bibfield  {author} {\bibinfo {author} {\bibfnamefont {A.}~\bibnamefont
  {{Passamonti}}}, \bibinfo {author} {\bibfnamefont {N.}~\bibnamefont
  {{Andersson}}},\ and\ \bibinfo {author} {\bibfnamefont {W.~C.~G.}\
  \bibnamefont {{Ho}}},\ }\href {https://doi.org/10.1093/mnras/stv2149}
  {\bibfield  {journal} {\bibinfo  {journal} {\mnras}\ }\textbf {\bibinfo
  {volume} {455}},\ \bibinfo {pages} {1489} (\bibinfo {year}
  {2016})}\BibitemShut {NoStop}%
\bibitem [{\citenamefont {{Flanagan}}\ and\ \citenamefont
  {{Racine}}(2007)}]{flanaganracine07}%
  \BibitemOpen
  \bibfield  {author} {\bibinfo {author} {\bibfnamefont {{\'E}.~{\'E}.}\
  \bibnamefont {{Flanagan}}}\ and\ \bibinfo {author} {\bibfnamefont
  {{\'E}.}~\bibnamefont {{Racine}}},\ }\href
  {https://doi.org/10.1103/PhysRevD.75.044001} {\bibfield  {journal} {\bibinfo
  {journal} {\prd}\ }\textbf {\bibinfo {volume} {75}},\ \bibinfo {eid} {044001}
  (\bibinfo {year} {2007})}\BibitemShut {NoStop}%
\bibitem [{\citenamefont {{Poisson}}(2020)}]{poisson20}%
  \BibitemOpen
  \bibfield  {author} {\bibinfo {author} {\bibfnamefont {E.}~\bibnamefont
  {{Poisson}}},\ }\href {https://doi.org/10.1103/PhysRevD.101.104028}
  {\bibfield  {journal} {\bibinfo  {journal} {\prd}\ }\textbf {\bibinfo
  {volume} {101}},\ \bibinfo {eid} {104028} (\bibinfo {year}
  {2020})}\BibitemShut {NoStop}%
\bibitem [{\citenamefont {{Xu}}\ and\ \citenamefont {{Lai}}(2017)}]{xulai17}%
  \BibitemOpen
  \bibfield  {author} {\bibinfo {author} {\bibfnamefont {W.}~\bibnamefont
  {{Xu}}}\ and\ \bibinfo {author} {\bibfnamefont {D.}~\bibnamefont {{Lai}}},\
  }\href {https://doi.org/10.1103/PhysRevD.96.08300510.48550/arXiv.1708.01839}
  {\bibfield  {journal} {\bibinfo  {journal} {\prd}\ }\textbf {\bibinfo
  {volume} {96}},\ \bibinfo {eid} {083005} (\bibinfo {year}
  {2017})}\BibitemShut {NoStop}%
\bibitem [{\citenamefont {{Kuan}}\ \emph {et~al.}(2021)\citenamefont {{Kuan}},
  \citenamefont {{Suvorov}},\ and\ \citenamefont {{Kokkotas}}}]{kuanetal21}%
  \BibitemOpen
  \bibfield  {author} {\bibinfo {author} {\bibfnamefont {H.-J.}\ \bibnamefont
  {{Kuan}}}, \bibinfo {author} {\bibfnamefont {A.~G.}\ \bibnamefont
  {{Suvorov}}},\ and\ \bibinfo {author} {\bibfnamefont {K.~D.}\ \bibnamefont
  {{Kokkotas}}},\ }\href {https://doi.org/10.1093/mnras/stab1898} {\bibfield
  {journal} {\bibinfo  {journal} {\mnras}\ }\textbf {\bibinfo {volume} {506}},\
  \bibinfo {pages} {2985} (\bibinfo {year} {2021})}\BibitemShut {NoStop}%
\bibitem [{\citenamefont {{Kuan}}\ \emph {et~al.}(2022)\citenamefont {{Kuan}},
  \citenamefont {{Kr{\"u}ger}}, \citenamefont {{Suvorov}},\ and\ \citenamefont
  {{Kokkotas}}}]{kuanetal22}%
  \BibitemOpen
  \bibfield  {author} {\bibinfo {author} {\bibfnamefont {H.-J.}\ \bibnamefont
  {{Kuan}}}, \bibinfo {author} {\bibfnamefont {C.~J.}\ \bibnamefont
  {{Kr{\"u}ger}}}, \bibinfo {author} {\bibfnamefont {A.~G.}\ \bibnamefont
  {{Suvorov}}},\ and\ \bibinfo {author} {\bibfnamefont {K.~D.}\ \bibnamefont
  {{Kokkotas}}},\ }\href
  {https://doi.org/10.1093/mnras/stac110110.48550/arXiv.2204.08492} {\bibfield
  {journal} {\bibinfo  {journal} {\mnras}\ }\textbf {\bibinfo {volume} {513}},\
  \bibinfo {pages} {4045} (\bibinfo {year} {2022})}\BibitemShut {NoStop}%
\bibitem [{\citenamefont {{Rau}}\ and\ \citenamefont
  {{Wasserman}}(2018)}]{rauwasserman18}%
  \BibitemOpen
  \bibfield  {author} {\bibinfo {author} {\bibfnamefont {P.~B.}\ \bibnamefont
  {{Rau}}}\ and\ \bibinfo {author} {\bibfnamefont {I.}~\bibnamefont
  {{Wasserman}}},\ }\href {https://doi.org/10.1093/mnras/sty2458} {\bibfield
  {journal} {\bibinfo  {journal} {\mnras}\ }\textbf {\bibinfo {volume} {481}},\
  \bibinfo {pages} {4427} (\bibinfo {year} {2018})}\BibitemShut {NoStop}%
\bibitem [{\citenamefont {{Agathos}}\ \emph {et~al.}(2015)\citenamefont
  {{Agathos}}, \citenamefont {{Meidam}}, \citenamefont {{Del Pozzo}},
  \citenamefont {{Li}}, \citenamefont {{Tompitak}}, \citenamefont {{Veitch}},
  \citenamefont {{Vitale}},\ and\ \citenamefont {{Van Den
  Broeck}}}]{agathosetal15}%
  \BibitemOpen
  \bibfield  {author} {\bibinfo {author} {\bibfnamefont {M.}~\bibnamefont
  {{Agathos}}}, \bibinfo {author} {\bibfnamefont {J.}~\bibnamefont {{Meidam}}},
  \bibinfo {author} {\bibfnamefont {W.}~\bibnamefont {{Del Pozzo}}}, \bibinfo
  {author} {\bibfnamefont {T.~G.~F.}\ \bibnamefont {{Li}}}, \bibinfo {author}
  {\bibfnamefont {M.}~\bibnamefont {{Tompitak}}}, \bibinfo {author}
  {\bibfnamefont {J.}~\bibnamefont {{Veitch}}}, \bibinfo {author}
  {\bibfnamefont {S.}~\bibnamefont {{Vitale}}},\ and\ \bibinfo {author}
  {\bibfnamefont {C.}~\bibnamefont {{Van Den Broeck}}},\ }\href
  {https://doi.org/10.1103/PhysRevD.92.023012} {\bibfield  {journal} {\bibinfo
  {journal} {\prd}\ }\textbf {\bibinfo {volume} {92}},\ \bibinfo {eid} {023012}
  (\bibinfo {year} {2015})}\BibitemShut {NoStop}%
\bibitem [{\citenamefont {{Andersson}}\ and\ \citenamefont
  {{Kokkotas}}(2001)}]{anderssonkokkotas01}%
  \BibitemOpen
  \bibfield  {author} {\bibinfo {author} {\bibfnamefont {N.}~\bibnamefont
  {{Andersson}}}\ and\ \bibinfo {author} {\bibfnamefont {K.~D.}\ \bibnamefont
  {{Kokkotas}}},\ }\href {https://doi.org/10.1142/S0218271801001062} {\bibfield
   {journal} {\bibinfo  {journal} {International Journal of Modern Physics D}\
  }\textbf {\bibinfo {volume} {10}},\ \bibinfo {pages} {381} (\bibinfo {year}
  {2001})}\BibitemShut {NoStop}%
\bibitem [{\citenamefont {{Lockitch}}\ \emph {et~al.}(2003)\citenamefont
  {{Lockitch}}, \citenamefont {{Friedman}},\ and\ \citenamefont
  {{Andersson}}}]{lockitchetal03}%
  \BibitemOpen
  \bibfield  {author} {\bibinfo {author} {\bibfnamefont {K.~H.}\ \bibnamefont
  {{Lockitch}}}, \bibinfo {author} {\bibfnamefont {J.~L.}\ \bibnamefont
  {{Friedman}}},\ and\ \bibinfo {author} {\bibfnamefont {N.}~\bibnamefont
  {{Andersson}}},\ }\href {https://doi.org/10.1103/PhysRevD.68.124010}
  {\bibfield  {journal} {\bibinfo  {journal} {\prd}\ }\textbf {\bibinfo
  {volume} {68}},\ \bibinfo {eid} {124010} (\bibinfo {year}
  {2003})}\BibitemShut {NoStop}%
\bibitem [{\citenamefont {{Idrisy}}\ \emph {et~al.}(2015)\citenamefont
  {{Idrisy}}, \citenamefont {{Owen}},\ and\ \citenamefont
  {{Jones}}}]{idrisyetal15}%
  \BibitemOpen
  \bibfield  {author} {\bibinfo {author} {\bibfnamefont {A.}~\bibnamefont
  {{Idrisy}}}, \bibinfo {author} {\bibfnamefont {B.~J.}\ \bibnamefont
  {{Owen}}},\ and\ \bibinfo {author} {\bibfnamefont {D.~I.}\ \bibnamefont
  {{Jones}}},\ }\href {https://doi.org/10.1103/PhysRevD.91.024001} {\bibfield
  {journal} {\bibinfo  {journal} {\prd}\ }\textbf {\bibinfo {volume} {91}},\
  \bibinfo {eid} {024001} (\bibinfo {year} {2015})}\BibitemShut {NoStop}%
\end{thebibliography}%

\end{document}